\documentclass[prl,twocolumn,superscriptaddress,twocolumn]{revtex4}
\usepackage{bm}
\usepackage{amssymb,xcolor}
\usepackage{dcolumn}
\usepackage{bm}
\usepackage{graphicx}
\usepackage{amsmath}
\usepackage{latexsym}
\usepackage{amsfonts}
\usepackage{amssymb}
\usepackage{array}
\usepackage{epsfig}
\usepackage{epstopdf}
\usepackage{times}
\usepackage{amsmath,epsfig}

\newcommand{\ket}[1]{\left\vert#1\right\rangle}

\newcommand{\bra}[1]{\left\langle#1\right\vert}

\newcommand{\id}{\leavevmode\hbox{\small1\normalsize\kern-.33em1}}
\newcommand{\beq}{\begin{equation}}
\newcommand{\eeq}{\end{equation}}

\begin{document}

\title{Linear Optics  Simulation of Non-Markovian Quantum Dynamics}
\author{Andrea Chiuri}
\email{a.chiuri@uniroma1.it}
\affiliation{Dipartimento di Fisica, Sapienza Universit\`a di Roma, Piazzale Aldo Moro 5, I-00185 Roma, Italy}
\author{Chiara Greganti}
\affiliation{Dipartimento di Fisica, Sapienza Universit\`a di Roma, Piazzale Aldo Moro 5, I-00185 Roma, Italy}
\author{Laura Mazzola}
\email{l.mazzola@qub.ac.uk}
\affiliation{Centre for Theoretical Atomic, Molecular and Optical Physics, School of Mathematics and Physics, Queen's University Belfast, BT7 1NN Belfast, United Kingdom}
\author{Mauro Paternostro}
\affiliation{Centre for Theoretical Atomic, Molecular and Optical Physics, School of Mathematics and Physics, Queen's University Belfast, BT7 1NN Belfast, United Kingdom}
\affiliation{Institut f\"ur Theoretische Physik, Albert-Einstein-Allee 11, Universit\"at Ulm, D-89069 Ulm, Germany}
\author{Paolo Mataloni}
\affiliation{Dipartimento di Fisica, Sapienza Universit\`a di Roma, Piazzale Aldo Moro 5, I-00185 Roma, Italy}
\affiliation{Istituto Nazionale di Ottica (INO-CNR), Largo E. Fermi 6, I-50125 Firenze, Italy}
\maketitle

{The simulation of quantum processes is a key goal for the grand programme aiming at grounding quantum technologies as the way to explore complex phenomena that are 
inaccessible through standard, classical calculators~\cite{BulutaSci09,walther}. Some interesting steps have been performed in this direction: simple condensed matter 
and chemical processes have been implemented on controllable quantum simulators~\cite{WaltherHeisenberg, WhiteHydrogen}. The relativistic motion and scattering of a 
particle in the presence of a linear potential has been demonstrated in a trapped-ion quantum simulator~\cite{GerritsmaNature, GerritsmaPRL,BlattNat12} that opens 
up the possibility to the study of quantum field theories~\cite{CasanovaPRL,CasanovaPRL1}. The quantum Ising model~\cite{sachdev} has been experimentally analysed 
under the perspective of universal digital quantum simulators~\cite{LanyonScience} and very recently scaled up to  hundreds of particles~\cite{Britton}. This scenario has recently been extended to open quantum evolutions~\cite{Barreiro}, marking the possibility to investigate important 
features of the way a quantum system interacts with its environment, including the so-called sudden death of entanglement induced by a memoryless environment~\cite{alme07sci}.

This interaction can destroy the most genuine quantum properties of the system, or involve exchange of coherence between system and environment, giving 
rise to memory effects and thus making the dynamics ``non-Markovian"~\cite{BreuerPetruccione}. Characterizing non-Markovian evolutions is currently at the 
centre of extensive theoretical and experimental efforts~\cite{guo,Apollaro}. Here we demonstrate experimentally the (non-)Markovianity of a process where 
system and environment are coupled through a simulated transverse Ising model~\cite{sachdev} (see Appendix). By engineering the evolution 
in a fully controlled photonic quantum simulator, we assess and demonstrate the role that system-environment correlations have in the emergence of memory effects.}

The paradigmatic description of quantum open dynamics involves a physical system $S$ evolving freely according to the Hamiltonian $H^S$ 
and embedded in an environment $E$ (whose free dynamics is ruled by $H^E$). System and environment interact via the Hamiltonian $H^{SE}$~\cite{BreuerPetruccione}, 
which we assume to be time-independent for easiness of description. While the dynamics of the joint state $\rho_{SE}$ is closed and governed by the unitary 
operator $U_t=e^{-\imath (H^S+H^E+H^{SE})t}$, the state of $S$, which is typically the only object to be accessible directly, is given by the reduced density 
matrix $\rho_S(t)=\mathrm{Tr}_E\{{\rho_{SE}(t)}\}$. For factorized initial states $\rho_{SE}(0)=\rho_S(0)\otimes\rho_E(0)$ and moving to an interaction frame 
defined by the free Hamiltonian of the total system, such reduced evolution can be recast in the operator-sum picture $\rho_S(t)=\sum_{\mu}K_\mu(t)\rho_S(0)K^\dag_\mu$, where $\{K_\mu\}$ 
is the set of trace-preseving, non-unitary Kraus operators of $S$ that are responsible for effects such as the loss of populations and coherence from the state of the system~\cite{Nielsen}. 

The roots of such {\it decoherence} mechanism have long been studied, together with their intimate connection to the so-called measurement problem and the implications of the collapse of 
the wave function~\cite{ZurekPT,ZurekRev,Schlosshauer}. Numerous experimentally oriented techniques have been proposed to counteract decoherence~\cite{ZanardiPRL97,Nielsen,ViolaPRA98}. 
Yet, a complementary viewpoint can be taken, where the possibility to engineering structured environments and tailored $S$-$E$ couplings is seen as a resource to achieve longer coherence 
times of the system~\cite{Myatt}, prepare entangled states, perform quantum computation, and realise quantum memories. This calls for the exploitation of memory-effects typical of a 
non-Markovian dynamics as a useful tool for the processing of a quantum state.  

Unfortunately, a satisfactory understanding of non-Markovianity is yet to be reached, which motivates the recent and intense efforts performed towards  the rigorous characterisation of non-Markovian evolutions, the formulation of criteria for the emergence of non-Markovian features, and the proposition of 
factual measures for the quantification of the degree of non-Markovianity of a process ~\cite{Eisert,Rivas,Breuer,Sun}. Some of them have been recently used in order to characterize, 
both theoretically and experimentally, the character of system-environment interactions for few- and many-body quantum systems~\cite{guo,Apollaro}.

\begin{figure}[t]
\includegraphics[width=\linewidth]{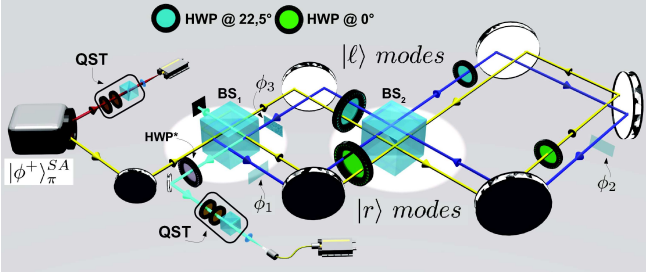}
\caption{
{\bf The Sagnac interferometer used to perform our simulation}. The red line represents the high photon, whose polarisation is immediately detected. The low photon, depicted by the yellow and blue paths, goes through the set of gates acting on its polarisation and momentum, implemented by BS, HWP, and glasses plates $\phi_i$. The optical axis of $\mathrm{HWP}^*$ is kept free so as to implement the last step of the simulation. The thin glass plates $\phi_1$, $\phi_2$ and $\phi_3$ placed in the interferometer allow to set the phase of the environment evolution at each step. The Hadamard gates are implemented by setting $\phi_1=\phi_2=\phi_3=0$ [cf. Eq.~(\ref{evol})]. } 
\label{setup}
\end{figure}

Here we use a photonic setup to simulate a system-environment coupling ruled by a transverse Ising spin model (see Appendix). 
Such interaction gives rise to the non-Markovian evolution of $S$, as witnessed by the non-monotonic behavior of its entanglement with an ancilla $A$ 
that is shielded from the environmental effects. Our goal is the experimental investigation of the fundamental connection between system-environment 
correlations and non-Markovianity.
The simulator that we propose allows for the implementation of various free evolutions of $S$ and $E$, as well as the adjustment of their mutual coupling, 
thus making possible the transition from deeply non-Markovian dynamics, all the way down to a fully forgetful regime. It is realised using different degrees 
of freedom of the photonic device shown in Fig.~\ref{setup}, which consists of the concatenation of two Mach-Zehnder interferometers and a single Sagnac loop. 
The information carriers are two photons (referred hereafter as ``high" and ``low"): the system $S$ is embodied by the polarisation of the low photon (which 
can be either horizontal $\ket{H}\equiv\ket{0}$ or vertical $\ket{V}\equiv\ket{1}$), while the environment $E$ is encoded in the longitudinal momentum degree 
of freedom (the path) of the same photon (which will be right $\ket{r}\equiv\ket{0}$ or left $\ket{l}\equiv\ket{0}$). The ancilla $A$ is embodied by the polarisation of the high photon. High and low photons are emitted by the source of polarization-entangled described in Appendix. 

The ancilla is the key tool for our goals. Indeed, to investigate the emergence of non-Markovianity in the evolution of $S$ due to its interaction with $E$, 
we use the method proposed in Ref.~\cite{Rivas}: we focus on the modifications induced by the $S$-$E$ coupling on a prepared entangled state of $S$ and $A$. 
If the $S$-$A$ entanglement decays monotonically in time, the dynamics of $S$ is fully Markovian. Differently, if for certain time-windows there 
is a kick-back from $E$ that makes such entanglement increase, the dynamics is necessarily non-Markovian. In fact, if the local action of the environment is 
no longer represented by a continuous family of completely positive maps, the $S$-$A$ entanglement is no longer constrained to decrease monotonically. 
This is evidence of the flow-back of coherence on the system and results in an increment of the $S$-$A$ entanglement.

\begin{figure}[t]
\begin{center}
\includegraphics[width= \linewidth]{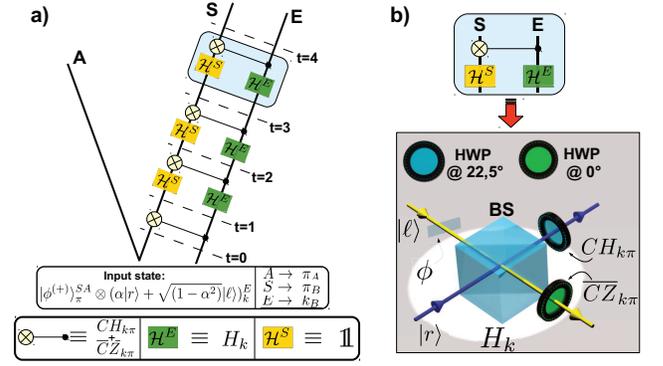}
\caption{{\bf Scheme of principle of the simulation}. 
a) A maximally entangled state of $A$ and $S$ is prepared. $S$ and $E$ then evolve through the application of free single-qubit gates ($H^E$ and $H^S$) and joint two-qubits 
($C{\cal H}_{k\pi} \overline{CZ}_{k\pi}$) ones. Time increases from bottom to top of the figure. Here $\pi_h (\ell)$ stands for the polarization of the high (low) photon, while $k_\ell$ for the momentum of the low photon. b) Each block of the scheme is technically realised by the concatenation of a BS and two HWPs oriented at different angles.} 
\label{scheme}
\end{center}
\end{figure}

As in other digital quantum simulators, the dynamics is approximated by a stroboscopic sequence of quantum gates. Conceptually, the simulation consists 
of the forward evolution of $S$ over discrete time slices~\cite{LloydScie96} according to a Trotter-Suzuki decomposition~\cite{Trotter} of the total 
time propagator (see Appendix). This approach is known to be effective for quantum simulation~\cite{BlattNat12} and is implemented here 
by the sequence of operations shown in Fig.~\ref{scheme}a). The free evolution of the environment is accounted for by the Hadamard gate 
${\cal H}^E=(\sigma^E_z+\sigma^E_x)/\sqrt{2}$ ($\sigma^j_m$ is the $m=x,y,z$ Pauli matrix of qubit $j=S,E$ and ${\cal H}^j$ is the Hadamard gate of qubit $j=S,E$), while $H^S\propto\sigma^S_z$. 
The $S$-$E$ interaction is engineered by implementing the controlled-rotation $G^{ES}=\ket{0}\bra{0}_E\otimes\openone^S+\ket{1}\bra{1}_E\otimes {\cal R}^S(\varphi)$, 
which rotates the system according to the general single-qubit operation ${\cal R}^S(\varphi)=\cos\varphi\openone^S-\imath\sin\varphi\sigma^S_y$ depending on 
the state of the environment. This class of conditional operations is obtained from Hamiltonian generators of the two-qubit transverse Ising 
form (see Appendix), which motivates our choice and is thus the class of system-environment interactions that is simulated in this work. 
Experimentally, we fix $\varphi=\pi/4$ so as to implement a conditional-($\sigma^S_x {\cal H}^S$) gate. By using the identity $\sigma^S_z={\cal H}^S\sigma^S_x{\cal H}^S$, 
it is straightforward to prove that this gate is locally equivalent (via $\sigma^S_z$) to the composition of a controlled Hadamard and a controlled anti-Z 
(i.e. a gate applying $\sigma^S_z$ only when $E$ is in $\ket{0}$), namely ${\cal G}^{ES}\equiv C{\cal H}_{k\pi} \overline{CZ}_{k\pi}=\ket{0}\bra{0}_E\otimes\sigma^S_z+\ket{1}\bra{1}_E\otimes {\cal H}^S$. 
The extra $\sigma^S_z$ needed to make the two conditional gates equivalent can thus be absorbed in $H^S$, which thus becomes $\openone^S$.

This set of gates is experimentally realised in our photonic simulator as sketched in Fig.\ref{scheme}b). The ${\cal H}^E$ gate is implemented by means 
of a beam splitter (BS) in conditions of temporal and spatial indistinguishability of the optical modes.
This scheme allows to evolve the input modes $\ket{r}$ and $\ket{\ell}$ at each step as
\begin{equation}\label{evol}
\begin{aligned}
\ket{r} \rightarrow\frac{1}{\sqrt{2}}(\ket{r} + e^{\imath\phi_i} \ket{\ell}),~~\ket{\ell} \rightarrow\frac{1}{\sqrt{2}}(\ket{r} - e^{\imath\phi_i} \ket{\ell}),
\end{aligned}
\end{equation}
where the phases $\phi_i$ are varied by rotating thin glass plates intercepting one of the optical modes entering the BS. In order to perform a more general rotation of $E$ it will be necessary to unbalance the output modes (using intensity attenuators) making the probability of occurrence of $\ket{r}$ and $\ket{\ell}$ unequal.
 
The controlled gate $C{\cal H}_{k\pi}$ is realised by placing a half-wave plate (HWP) on one of the two output modes with optical axis at $22.5^\circ$ with respect to the vertical direction. The temporal delay introduced by this waveplate is compensated 
by another HWP on the opposite output mode with optical axis set at $0^\circ$ with respect  to the vertical direction. This implements the $\overline{CZ}$ gate, as shown in Fig.~\ref{scheme}b). 
In order to assess non-Markovianity, ancilla and system are prepared in the maximally entangled state $\ket{\phi^+}^{SA}_\pi=(\ket{HH}+\ket{VV})/\sqrt{2}$. This two-qubit Bell state is engineered using the polarization entanglement source of Ref.~\cite{barb05pra} (see Appendix).  The environment is initialised in $\ket{\chi}_k^E=(\alpha \ket{r} +\sqrt{1-\alpha^2}\ket{\ell})^{E}_{k}$ ($\alpha\in\mathbb{R}$ is set by the transmittivity of $BS_1$ [cf. Fig.~\ref{setup}] and can be varied by placing an intensity attenuator on one of the two output modes), a state endowed with quantum coherence as it is key due to the conditional nature of the dynamics that we simulate.
The stability, modular structure and long coherence time of our interferometric setting  allows for the repeated iteration of each block of gates ${\cal G}^{ES}({\cal H}^E\otimes\openone^S)$.

As $A$ does not evolve with the environment, the polarisation state of  the high photon (depicted in red in Fig.~\ref{setup}) is immediately detected and only the low photon (in yellow in Fig.~\ref{setup}) goes through each sequence of gates at the various steps. 
A standard optical setup for the performance of quantum state tomography (QST)~\cite{jame01pra} is used to reconstruct the state of the $S$-$A$ system after the evolution. 
The radiation is collected by using an integrated system composed of GRaded INdex (GRIN) lens and single-mode fibre~\cite{ross09prl}, and is then detected by single-photon counters.
For each step of the simulated dynamics, we measure the state of the environment 
by projecting the polarisation qubits on the states $\ket{HH}$, $\ket{HV}$,
$\ket{VH}$, $\ket{VV}$. Finally, we trace out the degrees of freedom of the $S$-$A$ system by summing up the corresponding counts measured for every single projection 
needed for the implementation of single-qubit QST. The Pauli operators for $E$, generated after the first passage through the $BS_1$, are measured using $BS_2$. The same procedure is followed to perform the QST of the environmental state at each step. 
The state of the $S$-$A$ system is reconstructed in a similar way, by summing up the counts collected after projecting $E$ onto $\ket{r}$ and $\ket{\ell}$.

\begin{figure*}[bt]
\includegraphics[width=1\textwidth]{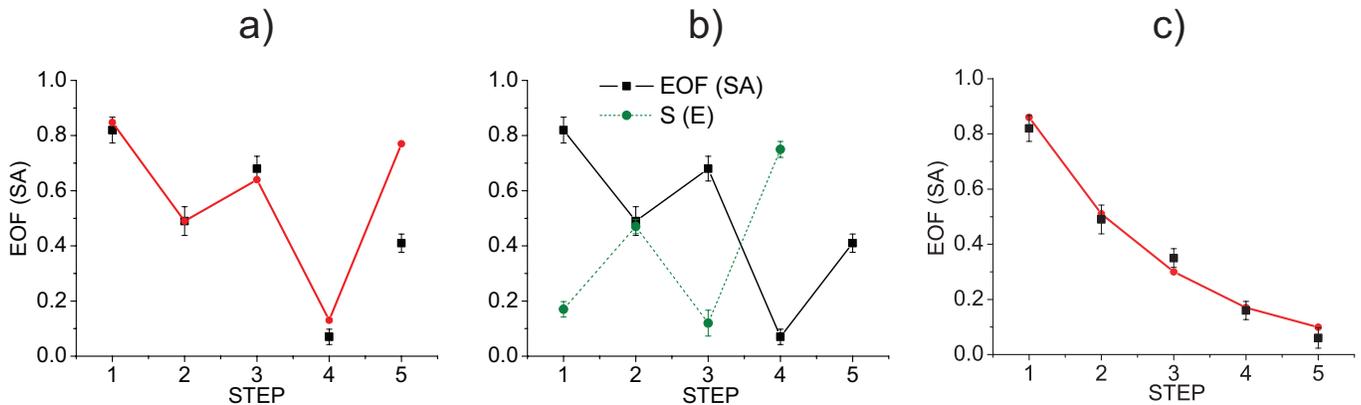}
\caption{{\bf Results of the simulation of the (non-)Markovian regimes}. a) Theoretical prediction (red line) against experimentally inferred data (black squares) for EOF(SA) as against the steps of the non-Markovian simulation. Two revivals of EOF(SA) are clearly visible. b) Comparison between the experimental evolution of EOF(SA) (black squares) and S(E) (green circles) representing between $E$ and $S$-$A$. The two evolutions are clearly anti-correlated. In order to measure the Pauli operators needed to reconstruct the state of $E$ at the fifth step, a further BS will be necessary. 
c) Theoretical behaviour (red line) and experimental data (black squares) for EOF(SA) in  the Markovian simulation.} 
\label{risultati}
\end{figure*}

Fig.~\ref{risultati} summarises the results obtained by running through the evolution of the overall system. In order to quantify entanglement we use the entanglement of 
formation EOF(SA)~\cite{Wootters} between $S$ and $A$, which is operatively linked to the {\it cost} of engineering a given state by means of Bell-state resources (see Appendix). 
We are also interested in the correlations shared by the environment with the rest of the system, 
hence we evaluate the von Neumann entropy of the environment, defined as $S(E)=-\mathrm{Tr}[\rho_E \log_2(\rho_E)]$, 
which under the assumption of pure total $ASE$ state quantifies the entanglement in the partition AS against E. 
Figure~\ref{risultati}a) shows the experimental entanglement of formation at each time step (black-square points) against the results of a theoretical model (red line) that, including all the most relevant sources of imperfections, 
deviates from the ideal picture sketched above. First, the BSs are not entirely polarisation-insensitive: for $BS_1$ and H (V) 
polarisation the reflectivity over transmittivity ratio $R/T=42/58$ (45/55), while  for $BS_2$ is $R/T=45/55$ (55/45). 
Second, although the desired input state $\ket{\phi^+}^{SA}_\pi$ is created with high fidelity ($\simeq93\%$, see Appendix), 
the entangled-state source generates spurious $\ket{HV}$ and $\ket{VH}$ components, accounting for about $5\%$ of the total state, 
which reduce the initial system-environment entanglement to about 0.8. The inclusion of such imperfections makes 
the agreement between theory and experimental data very good up to the fourth step of our simulation, 
showing at least one revival of the $S$-$A$ entanglement and thus witnessing the non-Markovian nature of the evolution~\cite{Rivas}. 
The fifth experimental point is significantly far from the theoretical behaviour because of the not-ideal setting of the phases 
$\phi_i$ ($i=1,2,3$). In fact, we have verified that their values are not 
completely polarization independent. This determines a slight difference for the four contributions
$\ket{HH}_{SA}$, $\ket{HV}_{SA}$,$\ket{VV}_{SA}$,$\ket{VH}_{SA}$ entering the state. This imperfection affects the performance of each Hadamard gate and becomes significant expecially for the last step, where
the cumulative effect of three ${\cal H}^E$ gates should be considered. Nonetheless, the last point reveals successfully the occurrence of a second entanglement revival, thus strengthening our conclusions.   

Figure~\ref{risultati}b) compares the experimentally inferred EOF(SA) (black squares) to the von Neumann entropy of the environment $S_E$ (green circles) quantifying the correlations shared between $E$ and $S$-$A$. These figures of merit appear to be perfectly anti-correlated, thus giving evidence of a trade-off between the amount of  entanglement that $S$ and $A$ can share at the expenses of $S$-$E$ correlations.  
This strengthens the idea that correlations with the environment play a fundamental role in this process, a point that has been addressed in Ref.~\cite{Mazzola} where it is shown that the establishment of system-environment correlations is a necessary conditions for the emergence of non-Markovianity. This result can be bridged with our analysis considering that the mixed system-environment state of Ref.~\cite{Mazzola} can be purified by enlarging the Hilbert space of the system including an appropriate ancilla.

To reinforce this point even further, we now explore the Markovian counterpart of  our simulation by replacing the unitary evolution of $E$ with an incoherent map that resets the environment into the very same state at each step of the evolution. As the key role in the $S$-$E$ interaction is played by quantum coherence, we chose to re-set the environment into a completely mixed state. This is realised by spoiling the temporal indistinguishability of the optical modes entering the BS by mutually delaying the $\ket{r}$ and $\ket{\ell}$ components. Intuitively, by making the state of the environment {\it rigid}, we wash out any possibility for system-environment correlations, thus pushing the dynamics towards Markovianity. This intuition is fully confirmed by the experimental evidences collected via QST: the black squares in Figure~\ref{risultati}c) show a monotonic (quasi-exponential) decay of EOF(SA) (matching our theoretical predictions, red line), which is in perfect agreement with the absence of $S$-$E$ quantum correlations as signalled by the positivity of the partially transposed $S$-$E$ state. 

By simulating a non-trivial two-qubit coupling model, we have demonstrated the non-Markovianity of the evolution induced on $S$ by a dynamical environment and a system-environment interaction 
allowing for kick-back of coherence. By adopting a witness that makes use of the effects that non-Markovianity has on entanglement, we have explored experimentally the link between the 
emergence of non-Markovianity and system-environment correlations. The next step in this endeavour will be the experimental proof that non-Markovianity can be used as a resource for the 
advantageous  processing of information, such as the preparation of interesting states, along the lines of previous studies on state engineering and information manipulation through 
Markovian processes~\cite{Kraus,Verstraete}. Our setup will be particularly well suited for this task, in light of the effective control over both system and environment that can be engineered both in space (thanks to the modular nature of the setting) and in time. 

\noindent
{\bf Acknowledgments} This work was supported by EU-Project CHISTERA-QUASAR, PRIN 2009 and FIRB-Futuro in ricerca HYTEQ, the EU under a Marie Curie IEF 
Fellowship (L.M.), and the UK EPSRC (M.P.) under a Career Acceleration Fellowship and a grant of the ``New Directions for Research Leaders" initiative (EP/G004579/1). 

\section{Appendix}

{\subsection{Source of entangled states} 
The black box shown in Fig.~1 of the main text represents the source 
used for the generation of the hyperentangled state encoded in the
polarization and path degrees of freedom of the photons produced by spontaneous parametric down-conversion. Here 
polarization entanglement, corresponding to the state $\ket{\phi^+}=\frac{1}{\sqrt{2}}(\ket{HH}+\ket{VV})$,  was realized by spatial and 
temporal superposition of photon pair emissions occurring with equal probability, back and forth, from a single barium 
borate (BBO) type-I crystal under double excitation of a vertically polarized UV CW pump beam. 
A suitable rotation of the polarization on one of the two possible 
emission directions of the photons was then applied. The source that we have used is such that the two photons, 
belonging to the degenerate BBO emission cone, are also path entangled, as explained in Ref.~\cite{barb05pra}. 
For our pourposes we need to select a pair of correlated directions, belonging to the emission cone surface,
along which two photons travel, obtaining in this way the high and low photons shown in red and yellow in Fig.~1.\\

\subsection{Entanglement of formation}
For two qubits, the entanglement of formation is given by~\cite{Wootters}
$\textrm{EOF}=h\left(\frac{1+\sqrt{1-C^2}}{2}\right)$,
with $h(x)=-x\log_2 x-(1-x) \log_2(1-x)$ and $C=\max\{0,\lambda_1-\lambda_2-\lambda_3-\lambda_4\}$. Here, $\{\lambda_i\}$ is the set of eigenvalues (arranged in non-increasing order) of the Hermitian matrix $R=\sqrt{\sqrt{\rho}\tilde{\rho}\sqrt{\rho}}$ 
where $\tilde{\rho}=(\sigma_y\otimes\sigma_y)\rho^*(\sigma_y\otimes\sigma_y)$ and $\rho^*$ is the conjugate of the density matrix}.

\subsection{Controlled-rotation gates from quantum Ising models}

Here we sketch the procedure for the achievement of a controlled rotation from a quantum Ising model and determine the conditions to 
achieve a desired rotation angle. For convenience of notation, we will use the correspondences (already introduced in the main text) 
$\{\ket{0} ,\ket{1} \}_{S,A}\rightarrow\{\ket{H},\ket{V}\}_{S,A}$ and $\{\ket{0} ,\ket{1} \}_{E} \rightarrow\{\ket{r},\ket{\ell}\}_E$. 

The transverse Ising-like model~\cite{sachdev} whose associated propagator has been simulated in our work is 
\beq
H^{Ising}=\epsilon^E_x\sigma^E_x+\sum_{p=x,y,z}\epsilon^S_p\sigma^S_p+J\sigma^E_z\sigma^S_z. 
\eeq
We take $\epsilon^E_x\gg J$, so that the dynamics of the $E$ subsystem as induced by this coupling is effectively frozen.   In the computational basis, this model can be written as
\beq
\begin{aligned}
H^{SE}&=
\ket{0}\bra{0}_E\otimes\left[
\begin{pmatrix}
\epsilon^S_z+J&\epsilon^S_x-\imath\epsilon^S_y\\
\epsilon^S_x+\imath\epsilon^S_y&-\epsilon^S_z-J
\end{pmatrix}
+\epsilon^E_z\openone^S\right]\\
&+
\ket{1}\bra{1}_E\otimes\left[
\begin{pmatrix}
\epsilon^S_z-J&\epsilon^S_x-\imath\epsilon^S_y\\
\epsilon^S_x+\imath\epsilon^S_y&-\epsilon^S_z+J
\end{pmatrix}
-\epsilon^E_z\openone^S\right].
\end{aligned}
\eeq 
The diagonal terms proportional to $\epsilon^E_z$ only give rise to a shift in the energy of the two subspaces that have been identified by this formal splitting. We can thus safely neglect them. On the other hand, each non-diagonal matrix acting in the $S$ subspace gives rise to a non-trivial single-qubit evolution of the form  
\beq
\begin{aligned}
{\cal U}^S_0&{=}
\begin{pmatrix}
\cos(\nu_0 t)-\frac{\imath(\epsilon^S_z+J)}{\nu_0}\sin(\nu_0t)&-\frac{\epsilon^S_y+\imath\epsilon^S_x}{\nu_0}\sin(\nu_0t)\\
\frac{\epsilon^S_y-\imath\epsilon^S_x}{\nu_0}\sin(\nu_0t)&\cos(\nu_0 t)+\frac{\imath(\epsilon^S_z+J)}{\nu_0}\sin(\nu_0t)
\end{pmatrix},\\
{\cal U}^S_1&{=}
\begin{pmatrix}
\cos(\nu_1 t)-\frac{\imath(\epsilon^S_z-J)}{\nu_1}\sin(\nu_1t)&-\frac{\epsilon^S_y+\imath\epsilon^S_x}{\nu_1}\sin(\nu_1t)\\
\frac{\epsilon^S_y-\imath\epsilon^S_x}{\nu_1}\sin(\nu_1t)&\cos(\nu_1t)+\frac{\imath(\epsilon^S_z-J)}{\nu_1}\sin(\nu_1t)
\end{pmatrix},
\end{aligned}
\eeq
where $\nu_{k}=2\pi\sqrt{(\epsilon^{S}_x)^2+[\epsilon^S_z+(-1)^kJ]^2+(\epsilon^{S}_y)^2}~(k=0,1)$. Needless to say, we want ${\cal U}^S_0=\openone^S$ for a time $\tau$ which the conditional gate is performed. On the other hand, we should enforce ${\cal U}^S_1={\cal R}^S(\varphi)$ at $\tau$. The first condition is straightforwardly fulfilled for $\nu_0\tau/(2\pi)=n\in\mathbb{Z}$. As for the second condition, using the definition of ${\cal R}^S(\varphi)$ given in the main text, we should enforce  
\beq
\begin{aligned}
&\cos\varphi=\cos(\nu_1\tau),~~\sin\varphi=(\epsilon^S_y/\nu_1)\sin(\nu_1\tau),\\
&0=[(\epsilon^S_z-J)/\nu_1]\sin(\nu_1\tau)=-(\epsilon^S_x/\nu_1)\sin(\nu_1\tau).
\end{aligned}
\eeq
These can be solved, at a chosen value of $J$, for $\epsilon^S_z=J$, which give $\nu_0=2\pi\sqrt{\epsilon^S_y+4J^2}$ and $\nu_1=2\pi\epsilon^S_y$. In turn, these give us
\beq
\epsilon^S_y=\frac{\varphi}{2\pi\tau},~~\sqrt{4J^2\tau^2+\frac{\varphi^2}{4\pi^2}}=n.
\eeq
As the value of $J$ is determined by the experimental setting chosen to implement the gate, these conditions fully specify the values of the parameters in the Ising model needed in order to implement a controlled rotation gate with an arbitrary value of the angle $\varphi$.

\subsection{Experimental system-ancilla density matrices}
\begin{figure*}[t!!]
\includegraphics[width=1\textwidth]{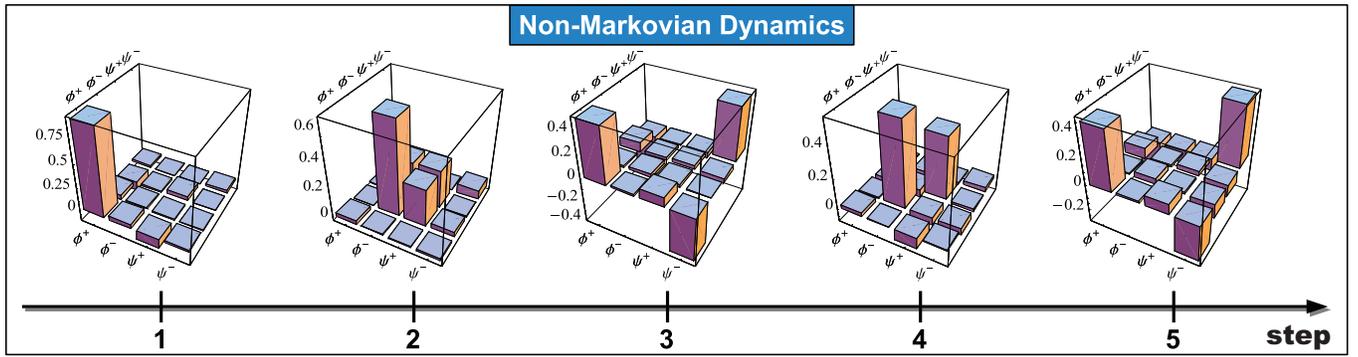}
\caption{Tomographic reconstruction of the state shared by $S$ and $A$ at each step of the Non-Markovian
dynamics. The initial state is represented by a $\ket{\phi^+}^{\pi}_{SA}$. It is worth noting that in the fourth step the coherence 
terms are null, in fact in this case we obtained the lowest value of $EOF_{SA}$. 
In the fifth step there is a revival of the coherence terms but with smaller values with respect to
the third step. The imaginary matrices are negligible for each reconstructed state.} 
\label{tomo}
\end{figure*}
In Fig.~\ref{tomo} we report the $S$-$A$ states that have been created experimentally at each step of the non-Markovian dynamics that we have simulated. We have used the operative approach put forward in Ref.~\cite{jame01pra}, which is the standard in the tomographic reconstruction of multi-qubit states, to determine the density matrices of the $S$-$A$ system, which have then been expressed in the Bell 
basis spanned by $\ket{\phi}^\pm=\frac{1}{\sqrt{2}}(\ket{00}\pm\ket{11})$ and $\ket{\psi}^\pm=\frac{1}{\sqrt{2}}(\ket{01}\pm\ket{10})$.


\begin{thebibliography}{9}

\bibitem{BulutaSci09}
Buluta, I., \& Nori, F. Quantum simulators. {\it Science} {\bf 326}, 108�111 (2009).

\bibitem{walther} Cirac, J. I., \& Zoller, P. Goals and opportunities in quantum simulation. {\it Nature Phys.} {\bf 8}, 264 (2012).

\bibitem{WaltherHeisenberg} Ma, X.-S. , Dakic, B.,  Naylor, W., Zeilinger, A., \& Walther, P. Quantum simulation of the wavefunction to probe frustrated Heisenberg spin systems. {\it Nature Phys.} {\bf }7 399 (2011).

\bibitem{WhiteHydrogen} Lanyon, B. P., {\it et al.} Towards quantum chemistry on a quantum computer. {\it Nature Chem.} {\bf 2}, 106 (2010).

\bibitem{GerritsmaNature} Gerritsma, R., {\it et al.} Quantum simulation of the Dirac equation. {\it Nature} {\bf 463}, 68 (2010).

\bibitem{GerritsmaPRL} Gerritsma, R., {\it et al.} Quantum Simulation of the Klein Paradox with Trapped Ions. {\it Phys. Rev. Lett.} {\bf 106}, 060503 (2011).

\bibitem{BlattNat12}
Blatt, R., \& Roos, C. F. Quantum simulations with trapped ions. {\it Nature Phys.} {\bf 8} 277 (2012).

\bibitem{CasanovaPRL} Casanova, J. {\it et al.} Quantum Simulation of Quantum Field Theories in Trapped Ions. {\it Phys. Rev. Lett.} {\bf 107}, 260501 (2011).

\bibitem{CasanovaPRL1} Casanova, J., Mezzacapo, A., Lamata, L., \& Solano, E. 	
Quantum Simulation of Interacting Fermion Lattice Models in Trapped Ions. {\it Phys. Rev. Lett.} {\bf 108}, 190502 (2012).

\bibitem{sachdev} Sachdev, S. {\it Quantum Phase Transitions} (Cambridge Univ. Press, 2011).

\bibitem{LanyonScience} Lanyon, B. P., {\it et al.} Universal Digital Quantum Simulation with Trapped Ions. {\it Science} {\bf 334}, 6052 (2011).

\bibitem{Britton} Britton, J. W., {\it et al.} Engineered two-dimensional Ising
interactions in a trapped-ion quantum simulator with hundreds of spins. {\it Nature} {\bf 484}, 489 (2012).

\bibitem{Barreiro}
Barreiro, J. T. {\it et al.} An open-system quantum simulator with trapped ions. {\it Nature} {\bf 470}, 486 (2011).

\bibitem{alme07sci}
Almeida, M. P., {\it et al.}. Environment-Induced Sudden Death of Entanglement. {\it Science} {\bf 316}, 579 (2007).

\bibitem{BreuerPetruccione}
Breuer, H-P. \& Petruccione, F. {\it The Theory of Open Quantum Systems} (Oxford Univ. Press, 2007).

\bibitem{guo}
Liu, B.-H., {\it et al.} Experimental control of the transition from Markovian to non-Markovian dynamics of open quantum systems {\it Nature Phys.} \textbf{7}, 931 (2011).

\bibitem{Apollaro} Apollaro, T. J. G., {\it et al.} Memory-keeping effects and forgetfulness in the dynamics of a qubit coupled to a spin chain. {\it Phys. Rev. A} {\bf 83}, 032103 (2011).

\bibitem{Nielsen}
Nielsen, M. A., \& Chuang, I. L. {\it Quantum Computation and Quantum Information} (Cambridge Univ. Press, 2000).

\bibitem{ZurekPT}
Zurek, W. H. Decoherence and the transition from quantum to classical. {\it Phys. Today} {\bf 44}, 36 (1991).

\bibitem{ZurekRev}
Zurek, W. H. Decoherence, einselection, and the quantum origins of the classical. {\it Rev. Mod. Phys.} {\bf 75}, 715 (2003).

\bibitem{Schlosshauer}
Schlosshauer, M. Decoherence, the measurement problem, and interpretations of quantum mechanics. {\it Rev. Mod. Phys.} {\bf 76}, 1267 (2005).

\bibitem{ZanardiPRL97}
Zanardi, P., \& Rasetti, M. Noiseless Quantum Codes. {\it Phys. Rev. Lett.} {\bf 79}, 3306 (1997).

\bibitem{ViolaPRA98}
Viola, L., \& Lloyd, S. Dynamical suppression of decoherence in two-state quantum systems. {\it Phys. Rev. A} {\bf 58}, 2733 (1998).

\bibitem{Myatt}
Myatt, C. J., {\it et al.} Decoherence of quantum superpositions through coupling to engineered reservoirs. {\it Nature} {\bf 403}, 269 (2000).

\bibitem{Eisert}
Wolf, M. M., Eisert, J., Cubitt, T. S., \& Cirac, J. I. Assessing Non-Markovian Quantum Dynamics. {\it Phys. Rev. Lett.} \textbf{101}, 150402 (2008).

\bibitem{Rivas}
Rivas, \'{A}., Huelga, S. F. \& Plenio, M. B. Entanglement and non-Markovianity of quantum evolutions. {\it Phys. Rev. Lett.} {\bf 105}, 050403 (2010).

\bibitem{Breuer}
Breuer, H-P., Laine, E-M., \& Piilo, J. Measure for the degree of non-Markovian behavior of quantum processes in open systems. {\it Phys. Rev. Lett.} {\bf 103}, 210401 (2009).

\bibitem{Sun} Lu, X.-M., Wang,  X., \& Sun, C. P. Quantum Fisher Information Flow in Non-Markovian Processes of Open Systems. {\it Phys. Rev. A} {\bf 82}, 042103 (2010). 

\bibitem{LloydScie96} Lloyd, S. Universal Quantum Simulators, {\it Science} {\bf 273}, 1073 (1996).

\bibitem{Trotter} Trotter, H. F. On the product of semi-groups of operators. {\it Proc. Am. Math. Soc.} {\bf 10}, 545 (1959). 

\bibitem{barb05pra}
Barbieri, M. {\it et al.} Polarization-momentum hyperentangled states: Realization and characterization. {\it Phys. Rev. A} {\bf 72}, 052110 (2005). 

\bibitem{jame01pra}
James, D., Kwiat, P., Munro, W. \& White, A. Measurement of qubits. {\it Phys. Rev. A} \textbf{64}, 052312 (2001).

\bibitem{ross09prl}
Rossi, A., Vallone, G., Chiuri, A., De Martini, F., \& Mataloni, P. Multipath entanglement of two photons. {\it Phys. Rev. Lett.} {\bf102}, 153902 (2009).

\bibitem{Wootters}
Wootters, W. K., {\it Phys. Rev. Lett.} {\bf 80}, 2245 (1998).

\bibitem{Mazzola}
Mazzola, L. Rodriguez-Rosario, C., Modi, K., \& Paternostro, M. Dynamical role of system-environment correlations in non-Markovian dynamics, arXiv:1203.3723 (2012).

\bibitem{Kraus} Kraus, B. {\it et al.} Preparation of Entangled States by Quantum Markov Processes. {\it Phys. Rev. A} {\bf 78}, 042307 (2008).

\bibitem{Verstraete} Vertraete, F., Wolf, M. M., \& Cirac, J. I. Quantum computation, quantum state engineering, and quantum phase transitions driven by dissipation. {\it Nature Phys.} {\bf 5}, 633 (2009).

\end{thebibliography}
\end{document}